\documentclass[11pt]{article}

\usepackage{amsmath,amsthm,amssymb,amscd}
\usepackage{geometry}
\usepackage[table]{xcolor}
\usepackage{graphicx}
\usepackage{subcaption}
\usepackage{multirow}
\usepackage{booktabs}
\usepackage{float}
\usepackage{tikz}
\usepackage{tikz-cd}
\usetikzlibrary{calc,positioning,3d}
\usepackage[normalem]{ulem}
\usepackage{array}
\usepackage{tabularx}
\usepackage{arydshln}
\usepackage{hyperref}
\usepackage{imakeidx}
\usepackage{makecell}
\usepackage{float}
\makeindex
\usepackage{seqsplit}

\usepackage{cleveref}
\usepackage{booktabs,array}
\usepackage{xr-hyper}
\usepackage{hyperref}
\usepackage{authblk}
\usepackage{lmodern}
\theoremstyle{definition}

\title{CANN: Commutative Algebra Neural Networks for Virus Classification}

\author[1]{Mushal Zia}
\author[2]{Faisal Suwayyid}
\author[1,3,4]{Guo-Wei Wei\thanks{
Corresponding author: Guo-Wei Wei
(\href{mailto:guowei.wei@uga.edu}{\texttt{guowei.wei@uga.edu}}).
}}

\affil[1]{Department of Mathematics, University of Georgia, Athens, GA 30602, USA}
\affil[2]{Department of Mathematics, King Fahd University of Petroleum and Minerals, Dhahran 31261, KSA}
\affil[3]{Department of Biochemistry and Molecular Biology, University of Georgia, Athens, GA 30602, USA}
\affil[4]{School of Computing, University of Georgia, Athens, GA 30602, USA}

\date{} 
\begin{document}
	\maketitle

\begin{abstract}
The extensive diversity of viral genomes challenges alignment-free representations to capture both nucleotide composition and the higher-order organization of recurring sequence patterns. Conventional $k$-mer representations quantify pattern abundance but do not explicitly encode how these patterns are organized across scales. Here, we introduce CANN, an alignment-free framework that combines persistent commutative-algebraic descriptors with neural representation learning for viral family classification. CANN uses persistent facet descriptors to encode the multiscale organization of $k$-mer patterns and complementary $k$-mer counts to capture sequence composition. Convolutional and Transformer encoders independently learn latent representations of these features, and cosine distances within their respective embedding spaces are averaged before $k$-nearest-neighbor classification. Across multiple NCBI viral genome collections, CANN achieves robust family-level performance, with the fusion of encoder-specific distances consistently improving classification relative to either encoder alone. Evaluation on genomes present in the 2026 NCBI collection but absent from the 2024 reference set further supports generalization to unseen members of families already represented in the reference data. Together, these findings show that persistent commutative-algebraic descriptors can complement neural representations of viral genomes, linking algebraic descriptions of genomic organization with data-driven classification.
\end{abstract}

\noindent\textbf{Keywords:} Viral family classification, Persistent commutative algebra, Stanley-Reisner theory, $k$-mer representations, Alignment-free sequence analysis, Convolutional neural networks, Transformers, Deep learning.

\newpage	
{\setcounter{tocdepth}{4} \tableofcontents}
\setcounter{page}{1}
\newpage

\section{Introduction}
\label{sec:introduction}

Accurate taxonomic classification of viruses is fundamental to organizing viral diversity, tracing evolutionary relationships, and interpreting the biological and ecological roles of viral communities \cite{murphy2012virus}. The rapid expansion of metagenomic sequencing has transformed virus discovery, revealing vast numbers of viral genomes that lack cultured representatives or experimental characterization and therefore must often be classified directly from sequence information  \cite{simmonds2017virus,simmonds2023four}. This expansion has created both an opportunity and a challenge for genome-based viral taxonomy. Unlike cellular organisms, which can often be classified using broadly conserved marker genes such as ribosomal RNA, viruses exhibit extraordinary diversity in genome type, length, segmentation, gene content, and genomic organization, and lack a universally conserved sequence marker \cite{simmonds2017virus}. Consequently, viral genomes can differ profoundly even among evolutionarily related groups, while distantly related viruses may share limited detectable sequence similarity. These features make robust classification across diverse viral genomes a challenging problem and motivate computational approaches that can capture the heterogeneous and evolutionarily complex nature of viral sequence diversity.

While alignment methods are crucial for identifying viral variants, alignment-based analyses remain central to identifying homologous sequences and reconstructing evolutionary relationships. Alignment-free methods provide a complementary approach by comparing sequence-derived representations without requiring explicit correspondence between sequence positions~\cite{zielezinski2017alignment}. Feature frequency profiles characterize genomes through normalized $k$-mer abundances~\cite{sims2009alignment}, whereas Markov-based composition approaches account for background expectations in word usage~\cite{qi2004whole}. Fourier power-spectrum methods summarize spectral properties of numerical sequence encodings~\cite{hoang2015new}, and natural-vector approaches combine composition with statistical moments of $k$-mer occurrence positions~\cite{sun2021geometric}. These representations capture different aspects of sequence variation, and large-scale benchmarking has shown that their relative performance depends on the biological task and dataset~\cite{zielezinski2019benchmarking}. Essentially, methods that capture the diversity of viral genomes often perform poorly in differentiating closely related viral families. Whereas methods that are able to detect subtle variants are unable to deal with a wide variety of viral genomes. This task dependence motivates the integration of complementary compositional and positional information.

Viral analysis tools also draw on different forms of genomic evidence. Gene-sharing approaches such as vConTACT2 and vConTACT3 use relationships among viral protein repertoires to support taxonomic assignment  \cite{bin2019taxonomic,bolduc2025machine}. PhaGCN integrates nucleotide features learned by convolutional neural networks with protein-similarity information in a graph-based framework for phage-contig classification~\cite{shang2021bacteriophage}. Notably,  geNomad combines gene-content information with a neural sequence classifier for virus and plasmid identification and uses marker proteins for viral taxonomic assignment~\cite{camargo2024identification}. VITAP uses protein-alignment evidence and hierarchy-aware taxonomic scoring to classify DNA and RNA viruses~\cite{zheng2025vitap}. For genomic prediction more broadly, DNABERT demonstrates how Transformer models can learn contextual representations of nucleotide sequences~\cite{ji2021dnabert}. VirSorter2 provides a multi-classifier, expert-guided approach to detect diverse viruses \cite{guo2021virsorter2}.

Multiscale mathematical representations offer another way to characterize genomic organization. Rather than considering only $k$-mer abundance, $k$-mer topology examines occurrence positions across filtration scales using topological and spectral information derived from persistent homology \cite{carlsson2009topology,papamarkou2024position}  and/or persistent Laplacians~\cite{hozumi2026revealing, su2025topological}. 
This approach stimulated the commutative algebra-based modeling of genomic sequences. Persistent Stanley-Reisner theory extends multiscale analysis to commutative algebra through rings, ideals, and associated invariants constructed from filtered simplicial complexes~\cite{suwayyid2025persistent}. Persistent commutative algebra has been successfully applied to the analysis of complex biological data \cite{feng2025caml, zia2027gbnl, zia2025cap, wee2025commutative}. By providing insights beyond those offered by conventional mathematical, statistical, and physical methods \cite{ren2025interpretability}, it has inspired related mathematical formulations \cite{hu2025commutative, suwayyid2025persistent, he2026localized} and is emerging as a promising paradigm in data science and machine learning \cite{wei2026commutative}.

Building on this theoretical and computational background, commutative algebra $k$-mer representations (CAKR) encode the positional organization of recurring sequence elements using persistent facet distances between genomes, with applications to genetic-variant classification, phylogenetic reconstruction, and viral family classification~\cite{suwayyid2026cakr}. These algebraic distances provide a mathematical basis for incorporating multiscale positional information into learned genomic representations. However, it is not clear whether a persistent commutative deep learning approach works for virus analysis.   

Here, we introduce CANN, an alignment-free framework that builds on CAKR by integrating persistent facet descriptors with neural representation learning for viral family classification. Whereas CAKR adopted a commutative algebra distance-based approach \cite{suwayyid2026cakr}, CANN uses commutative algebra to construct sequence descriptors for neural representation learning. Specifically, CANN combines multiscale positional information from persistent facet descriptors with complementary $k$-mer composition features. CNN and Transformer encoders independently learn embeddings from these features using known family labels. Within each embedding space, cosine distances are computed between test genomes and labelled reference genomes. The encoder-specific distances are then averaged to obtain a consensus distance for nearest-neighbour family assignment. This formulation brings together algebraic sequence descriptions, supervised representation learning, and distance fusion within a single classification framework.

We evaluate CANN for viral family classification on four benchmark collections derived from the National Center for Biotechnology Information (NCBI) viral genome resource: NCBI 2020, NCBI 2022, NCBI 2024, and NCBI 2024 All~\cite{brister2015ncbi}. Performance is assessed using 1-nearest-neighbour and 5-nearest-neighbour classification and reported alongside benchmark values available in the literature for six alignment-free methods: the natural vector method (NVM)~\cite{sun2021geometric}, the Markov $k$-string model (MKS)~\cite{qi2004whole}, feature frequency profiles~\cite{sims2009alignment,jun2010whole} with Jensen-Shannon divergence (FFP-JS) and Kullback-Leibler divergence (FFP-KL), the Fourier power-spectrum method (FPS)~\cite{hoang2015new}, and commutative algebra $k$-mer representations (CAKR)~\cite{suwayyid2026cakr}. Three additional evaluations using the 2026 NCBI collection examine the assignment of genomes absent from the 2024 reference set, cross-validation within the full 2026 collection, and assignment restricted to families represented in the 2024 reference set. 

Across the four benchmark collections, CANN achieves higher classification accuracy than the benchmark values reported for CAKR and the five other alignment-free methods considered. Combining CNN- and Transformer-derived cosine distances further improves classification relative to either encoder alone, supporting the use of complementary learned representations. Moreover, evaluation on previously unseen genomes from the 2026 NCBI collection supports generalization to additional members of families represented in the reference data. These findings establish CANN as an effective framework for viral family classification and demonstrate how multiscale algebraic descriptions of genomic organization can support neural representation learning.

\begin{figure}[htb!]
    \centering
 \includegraphics[width=\textwidth]{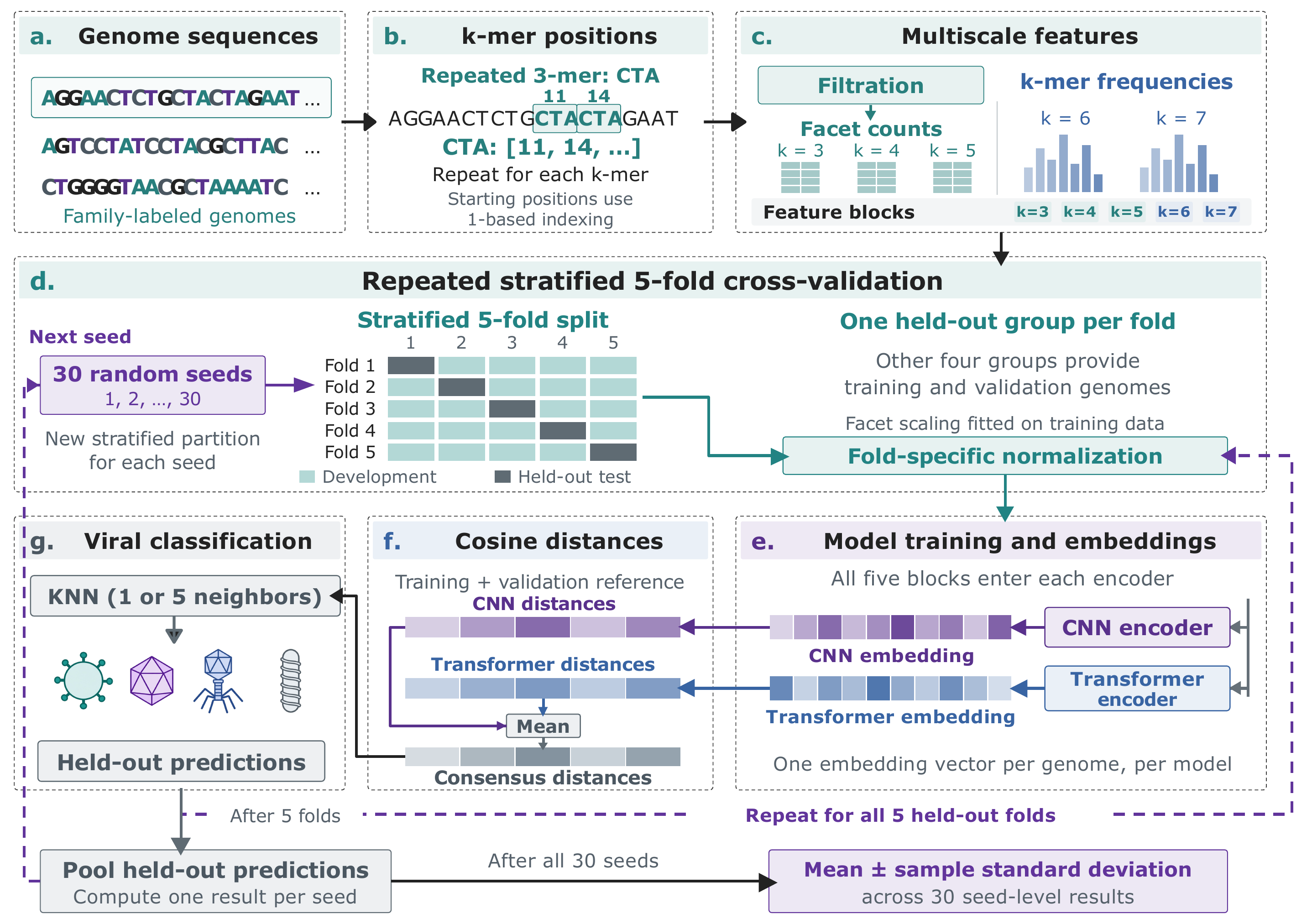}
    \caption{
    \textbf{CANN workflow for viral family classification.}
    (a-c) Family-labeled genomes are converted from $k$-mer occurrence positions into five feature blocks: filtration-based persistent facet-number features for $k=3,4,5$ and within-genome normalized $k$-mer frequency features for $k=6,7$.
    (d) Stratified five-fold cross-validation is repeated with a new partition for each of 30 random seeds, using fold-specific facet normalization fitted on the training data.
    (e-f) The unified five-block representation is used as input to CNN and Transformer encoders, which are initialized and trained independently within each cross-validation split. After checkpoint selection, cosine distances from held-out genomes to the common labeled training and validation reference set are computed in each encoder space and averaged with equal weights to produce consensus distances.
    (g) Each held-out genome is assigned to the family of its closest reference genome using consensus 1-NN, the primary CANN prediction. As an additional evaluation, consensus 5-NN assigns the genome to the family receiving the majority vote among its five closest reference genomes, with all five neighbors contributing equally. Predictions from the five held-out folds are pooled to obtain one performance result for each seed, and final results are reported as the mean $\pm$ sample standard deviation across the 30 seed-level results. Sequence examples and feature arrays are schematic.
    }
    \label{fig:flowchart}
\end{figure}

\section{Results and Discussion}

\subsection{An Overview of CANN}

Viral family classification benefits from representations that capture both sequence composition and the positional organization of genomic patterns. CANN addresses this need by combining commutative algebra descriptors, conventional $k$-mer frequency features, and supervised representation learning (Fig.~\ref{fig:flowchart}). Commutative algebra enters the framework through persistent facet-number descriptors, which encode the positional organization of $k$-mer occurrences.

A $k$-mer is a nucleotide word of length $k$, and its frequency provides information about sequence composition but does not describe the locations or spacing of its occurrences. CANN therefore records $k$-mer occurrence positions and uses filtration to construct persistent facet-number features for $k=3,4,5$. At each distance scale, a facet represents a maximal collection of linked occurrence positions; tracking these facets across scales captures the multiscale organization of repeated sequence patterns. These descriptors are evaluated at two filtration values, $0$ and $4^k$, yielding two scalar features for each possible $k$-mer. Conventional $k$-mer relative-frequency features for $k=6,7$ complement these positional descriptors. Together, these components form five input blocks containing 128, 512, 2,048, 4,096, and 16,384 scalar features for $k=3,4,5,6,7$, respectively, giving a total of 23,168 input features per genome.

The five feature blocks are assembled into a unified representation, which is used as an input to CNN and Transformer encoders. The two encoders are initialized and trained independently within each cross-validation split using viral family labels; no pretrained models are used. The CNN learns local combinations of features, whereas the Transformer’s self-attention mechanism captures relationships among feature components. Supervised training with viral family labels maps each genome to a compact embedding in each model space. Within each cross-validation fold, cosine distances between held-out genomes and the common labeled training and validation reference set are computed independently for the two encoders and averaged with equal weights to form consensus distances. These distances support 1-nearest-neighbor classification, the primary CANN prediction, and unweighted 5-nearest-neighbor majority voting as an additional evaluation. Stratified five-fold cross-validation is repeated with a new partition for each of 30 random seeds. Held-out predictions are pooled across the five folds for each seed, and performance is summarized as the mean $\pm$ sample standard deviation across the 30 seed-level results. This alignment-free workflow integrates sequence composition, multiscale positional organization, and complementary learned representations for viral family prediction.

\subsection{Viral Family Classification}
\label{sec:viral_family_classification}

CANN was evaluated on four viral-genome benchmark datasets derived from the NCBI Virus database: NCBI 2020~\cite{sun2021geometric}, NCBI 2022~\cite{yu2024optimal}, NCBI 2024~\cite{hozumi2026revealing}, and NCBI 2024 All~\cite{hozumi2026revealing}. Additional details regarding dataset construction, preprocessing, and family-size filtering are provided in Supplementary Section S1 and Supplementary Table S1.

\begin{figure}[htb!]
    \centering
    \includegraphics[width=0.9\textwidth]{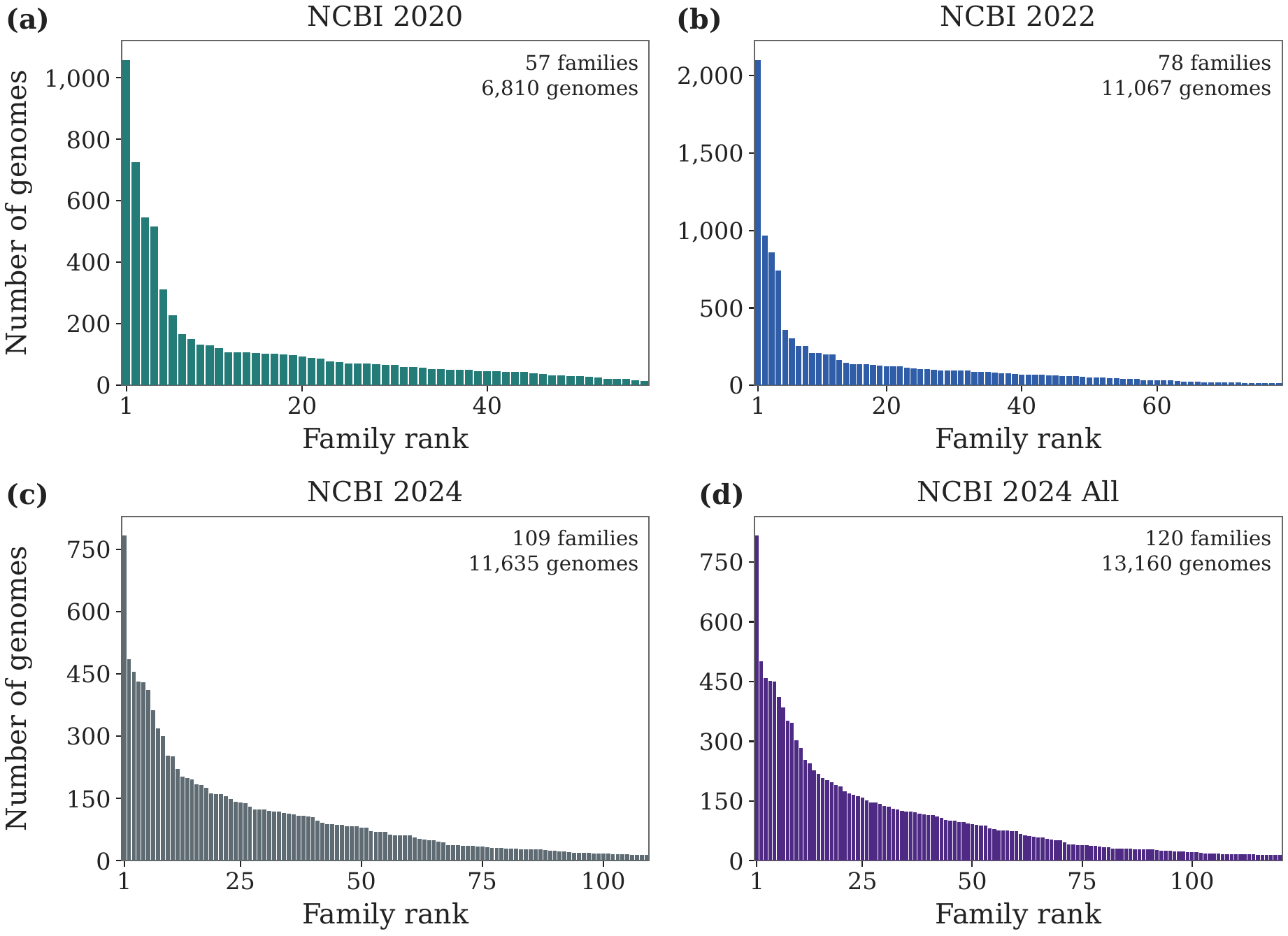}
    \caption{
    \textbf{Uneven viral-family representation across the four benchmark datasets.}
    Families are ranked by decreasing genome count after retaining only families represented by at least 15 genomes. The retained collections contain 6,810 genomes from 57 families in NCBI 2020, 11,067 genomes from 78 families in NCBI 2022, 11,635 genomes from 109 families in NCBI 2024, and 13,160 genomes from 120 families in NCBI 2024 All. The distributions remain strongly imbalanced in all four datasets.
    }
    \label{fig:family_distribution}
\end{figure}

A minimum family size of 15 genomes was applied to all four datasets. This retained 6,810 genomes from 57 families in NCBI 2020, 11,067 genomes from 78 families in NCBI 2022, 11,635 genomes from 109 families in NCBI 2024, and 13,160 genomes from 120 families in NCBI 2024 All. NCBI 2024 is a quality-controlled subset of NCBI 2024 All, whereas NCBI 2024 All retains the broader collection available from NCBI Virus. The corresponding family-size distributions are shown in Figure~\ref{fig:family_distribution}. All four datasets remain strongly imbalanced, with a few highly represented families and a long tail of smaller families.

The four datasets were evaluated using the same CANN configuration. We used repeated stratified five-fold cross-validation with 30 independent random seeds. Each seed generated a new fold partition. Normalization was fitted within each training fold. The validation data were used for checkpoint selection. The training and validation genomes then formed the reference collection for the held-out genomes. The CNN and Transformer encoders produced separate cosine-distance matrices, which were averaged with equal weights before nearest-neighbor classification. The closest reference genome determined the 1-NN prediction. The 5-NN prediction used an unweighted majority vote among the five closest reference genomes. Predictions from the five held-out folds were pooled within each seed. Thus, every genome contributed once to the seed-level evaluation. The main tables report mean values across the 30 seeds. Standard deviations and detailed encoder-specific results are provided in the Supplementary Section S4 and Table S2.

\begin{figure}[!htb]
    \centering
    \includegraphics[width=0.92\textwidth]{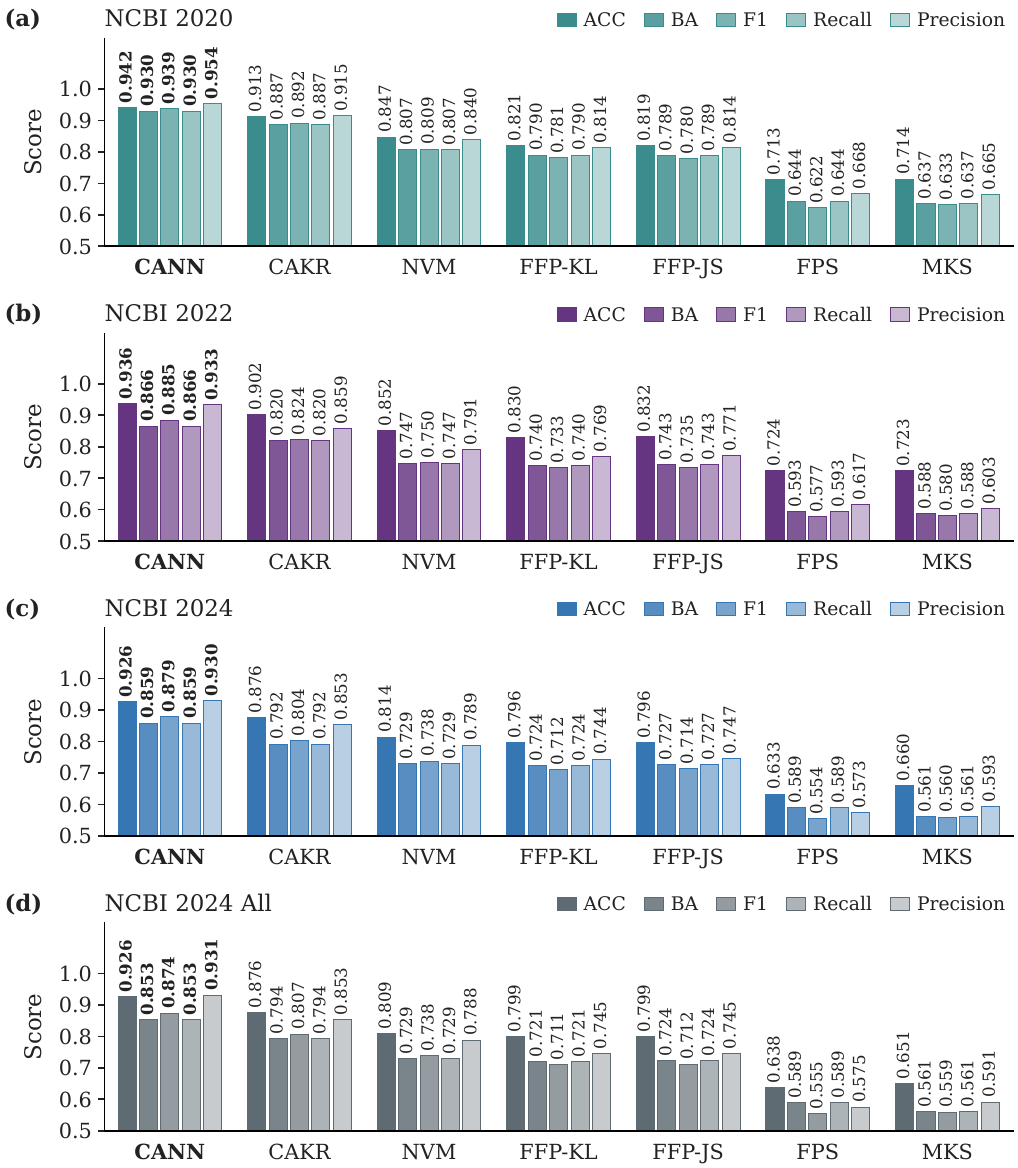}
    \caption{
    \textbf{Five-nearest-neighbor performance across four viral-genome datasets.}
    Results are shown for (a) NCBI 2020, (b) NCBI 2022, (c) NCBI 2024, and (d) NCBI 2024 All. Bars show accuracy, balanced accuracy, macro-F1, macro-recall, and macro-precision as proportions. All values use the 5-NN classification rule. CANN values represent means from repeated stratified five-fold cross-validation across 30 random seeds. Values for the comparison methods are the corresponding 5-NN benchmark values~\cite{hozumi2026revealing}. Standard deviations are provided for CANN in the Supplementary Table S2.
    }
    \label{fig:classification_results}
\end{figure}

\begin{table}[!htb]
\centering
\captionsetup{font=small}
\caption{\textbf{One-nearest-neighbor classification accuracy across four viral-genome datasets.}
CANN values are mean accuracies across 30 random-seed evaluations.
Comparison values are benchmark results available in the
literature~\cite{hozumi2026revealing,suwayyid2026cakr}. CANN standard deviations are reported
in the Supplementary Table S2. Boldface indicates the highest
accuracy within each dataset.}
\label{tab:cann_1nn_accuracy}
\fontsize{9.5}{11.5}\selectfont
\setlength{\tabcolsep}{3pt}
\renewcommand{\arraystretch}{1.08}
\begin{tabular}{@{}lccccccc@{}}
\toprule
Dataset & CANN & CAKR & NVM & FFP-JS & FFP-KL & FPS & MKS \\
\midrule
NCBI 2020
& \textbf{0.950} & 0.932 & 0.879 & 0.862 & 0.862 & 0.732 & 0.734 \\
NCBI 2022
& \textbf{0.946} & 0.920 & 0.875 & 0.870 & 0.870 & 0.732 & 0.735 \\
NCBI 2024
& \textbf{0.929} & 0.891 & 0.829 & 0.825 & 0.826 & 0.656 & 0.637 \\
NCBI 2024 All
& \textbf{0.929} & 0.892 & 0.825 & 0.832 & 0.832 & 0.647 & 0.647 \\
\bottomrule
\end{tabular}
\end{table}
Figure~\ref{fig:classification_results} provides the corresponding visual comparison. All plotted values use the 1-NN rule. CANN also showed the highest value for each displayed metric as shown in Table~\ref{tab:cann_1nn_accuracy}. The complete 5-NN comparison is also provided in Table~\ref{tab:cann_5nn_metrics}.

We first examined the encoder-derived distances within CANN. Under 1-NN classification, the CNN-derived, Transformer-derived, and combined CANN distances produced an accuracy of 92.40\%, 94.16\%, and 95.02\% on NCBI 2020, respectively. On NCBI 2022, the corresponding values were 91.81\%, 93.53\%, and 94.58\%. The values for NCBI 2024 were 88.16\%, 91.33\%, and 92.87\%. For NCBI 2024 All, they were 87.52\%, 91.19\%, and 92.94\%. The combined CANN distance produced the highest accuracy in all four datasets. These comparisons assess distance integration within CANN. They do not represent three independent classification methods.

Under 1-NN classification, our framework, CANN, achieved accuracies of 95.02\%, 94.58\%, 92.87\%, and 92.94\% on NCBI 2020, NCBI 2022, NCBI 2024, and NCBI 2024 All, respectively (Table~\ref{tab:cann_1nn_accuracy}). CANN yielded the highest reported accuracy for every dataset. Relative to CAKR, which achieved the highest benchmark accuracy among the comparison methods~\cite{suwayyid2026cakr}, the improvement ranged
from approximately 1.82 to 3.77 percentage points. The decrease in accuracy across the larger collections is consistent with their broader family coverage and more pronounced class imbalance.

The comparison was also extended to 5-NN classification to evaluate performance under the neighborhood size used by the published benchmark methods. CANN achieved accuracies of 94.25\%, 93.55\%, 92.57\%, and 92.60\% on NCBI 2020, NCBI 2022, NCBI 2024, and NCBI 2024 All, respectively as can be found in Table~\ref{tab:cann_5nn_metrics}. Moreover, our CANN framework also produced the highest balanced accuracy, macro-F1, and macro-precision within every dataset block. The accuracy advantage over CAKR, the strongest comparison method, ranged from approximately 2.95 to 5.00 percentage points under 5-NN. Furthermore, the consistently higher family-balanced scores indicate that the improvement was not restricted to the most abundant viral families. Although the family-balanced metrics declined as the datasets expanded, CANN maintained a clear advantage across both genome-level and family-level evaluation criteria. The individual CNN and Transformer results along with their standard deviations for both 1-NN and 5-NN are reported in the Supplementary Table S2.

\begin{table}[!htb]
\centering
\captionsetup{font=small}
\caption{\textbf{Five-nearest-neighbor classification performance across four viral-genome datasets.}
CANN values are means across 30 random-seed evaluations. Comparison
values are benchmark results available in the literature~\cite{hozumi2026revealing,suwayyid2026cakr}.
BA denotes balanced accuracy and equals macro-recall. F1 and precision
are macro-averaged. CANN standard deviations are reported in the
Supplementary Table S2. Boldface indicates the highest value within
each dataset block.}
\label{tab:cann_5nn_metrics}
\fontsize{9.5}{12}\selectfont
\setlength{\tabcolsep}{3.5pt}
\renewcommand{\arraystretch}{1.03}
\begin{tabular}{@{}llcccc@{}}
\toprule
Dataset & Method & ACC & BA & F1 & Prec. \\
\midrule
\multirow{7}{*}{NCBI 2020}
& \textbf{CANN} & \textbf{0.942} & \textbf{0.930} & \textbf{0.939} & \textbf{0.954} \\
& CAKR   & 0.913 & 0.887 & 0.892 & 0.915 \\
& NVM    & 0.847 & 0.807 & 0.809 & 0.840 \\
& FFP-KL & 0.821 & 0.790 & 0.781 & 0.814 \\
& FFP-JS & 0.819 & 0.789 & 0.780 & 0.814 \\
& MKS    & 0.714 & 0.637 & 0.633 & 0.665 \\
& FPS    & 0.713 & 0.644 & 0.622 & 0.668 \\
\midrule
\multirow{7}{*}{NCBI 2022}
& \textbf{CANN} & \textbf{0.936} & \textbf{0.866} & \textbf{0.885} & \textbf{0.933} \\
& CAKR   & 0.902 & 0.820 & 0.824 & 0.859 \\
& NVM    & 0.852 & 0.747 & 0.750 & 0.791 \\
& FFP-KL & 0.830 & 0.740 & 0.733 & 0.769 \\
& FFP-JS & 0.832 & 0.743 & 0.735 & 0.771 \\
& MKS    & 0.723 & 0.588 & 0.580 & 0.603 \\
& FPS    & 0.724 & 0.593 & 0.577 & 0.617 \\
\midrule
\multirow{7}{*}{NCBI 2024}
& \textbf{CANN} & \textbf{0.926} & \textbf{0.859} & \textbf{0.879} & \textbf{0.930} \\
& CAKR   & 0.876 & 0.792 & 0.804 & 0.853 \\
& NVM    & 0.814 & 0.729 & 0.738 & 0.789 \\
& FFP-KL & 0.796 & 0.724 & 0.712 & 0.744 \\
& FFP-JS & 0.796 & 0.727 & 0.714 & 0.747 \\
& MKS    & 0.660 & 0.561 & 0.560 & 0.593 \\
& FPS    & 0.633 & 0.589 & 0.554 & 0.573 \\
\midrule
\multirow{7}{*}{NCBI 2024 All}
& \textbf{CANN} & \textbf{0.926} & \textbf{0.853} & \textbf{0.874} & \textbf{0.931} \\
& CAKR   & 0.876 & 0.794 & 0.807 & 0.853 \\
& NVM    & 0.809 & 0.729 & 0.738 & 0.788 \\
& FFP-KL & 0.799 & 0.721 & 0.711 & 0.745 \\
& FFP-JS & 0.799 & 0.724 & 0.712 & 0.745 \\
& MKS    & 0.651 & 0.561 & 0.559 & 0.591 \\
& FPS    & 0.638 & 0.589 & 0.555 & 0.575 \\
\bottomrule
\end{tabular}
\end{table}

Figure~\ref{fig:family_distribution} shows a long tail of families represented by relatively few genomes across all four datasets. The larger collections include more viral families and exhibit lower family-balanced classification performance. For CANN, accuracy decreased by approximately 1.7 percentage points from NCBI 2020 to NCBI 2024 All, whereas balanced accuracy and macro-F1 decreased by approximately 7.7 and 6.5 percentage points, respectively. Thus, overall accuracy remained relatively stable, while performance averaged across families declined more substantially. The two 2024 collections showed a similar pattern: their accuracies were nearly identical, but NCBI 2024 All had lower balanced accuracy and macro-F1.

\subsubsection{Robustness of Method Rankings}
\label{sec:ranking_robustness}

We further examined whether the ordering of CANN, CAKR, NVM, FFP-KL, FFP-JS, FPS, and MKS remained consistent across evaluation metrics and benchmark datasets. Rankings were based on their 5-NN scores at the three-decimal precision used in the analysis, with rank 1 assigned to the best-performing method and average ranks assigned to ties. Agreement was measured using Kendall's coefficient of concordance, $W$, corrected for ties, where values close to 1 indicate strong agreement. The ranking procedure and mathematical definitions are provided in Supplementary Section S2.

Within each dataset, method rankings were compared across accuracy, balanced accuracy, macro-F1, macro-recall, and macro-precision. Concordance ranged from $W=0.978$ to $W=0.989$ (Table~\ref{tab:kendall_summary}). The strongest agreement occurred for NCBI 2022, while NCBI 2024 All had the lowest coefficient. Balanced accuracy and macro-recall were included separately, although they produced identical rankings and therefore did not provide distinct ranking information.

Rankings were also consistent across the four datasets for each metric. Concordance was $W=0.977$ for accuracy and macro-precision and $W=0.987$ for balanced accuracy, macro-F1, and macro-recall. CANN ranked first and CAKR ranked second under every metric in all four datasets, while minor changes occurred in the ordering of the remaining methods. These results demonstrate strong descriptive agreement in method ordering across the evaluated metrics and benchmark collections. Nominal $p$-values are provided for reference rather than treated as confirmatory evidence, given the small number of ranking lists and their dependence.

\begin{table}[!htb]
\centering
\captionsetup{font=small}
\caption{
\textbf{Agreement of method rankings across metrics and datasets.}
Kendall's coefficient of concordance, $W$, summarizes agreement among rankings of seven methods based on their 5-NN scores. Across-metric comparisons use accuracy, balanced accuracy, macro-F1, macro-recall, and macro-precision within each dataset. Across-dataset comparisons use the four benchmark collections for each metric. Scores are ranked at the three-decimal precision used in the analysis, with rank 1 assigned to the highest score and average ranks assigned to ties. Values of $W$ are corrected for ties. Balanced accuracy and macro-recall are retained separately, although they produce identical rankings. The statistic $\chi^2 = m(n-1)W$ uses $n=7$ methods, with $m=5$ for across-metric comparisons and $m=4$ for across-dataset comparisons. Nominal $p$-values use the chi-square approximation with six degrees of freedom and are provided for reference. Concordance is interpreted descriptively because the ranking lists are few and are not independent.
}
\label{tab:kendall_summary}

\fontsize{9.5}{12}\selectfont
\setlength{\tabcolsep}{4pt}
\renewcommand{\arraystretch}{1.08}

\begin{tabular}{@{}llccc@{}}
\toprule
\textbf{Comparison} &
\textbf{Dataset or metric} &
$\boldsymbol{W}$ &
$\boldsymbol{\chi^2}$ &
\textbf{$p$-value} \\
\midrule

\multirow{4}{*}{Across metrics}
& NCBI 2020     & 0.980 & 29.398 & 0.000051 \\
& NCBI 2022     & 0.989 & 29.657 & 0.000046 \\
& NCBI 2024     & 0.980 & 29.398 & 0.000051 \\
& NCBI 2024 All & 0.978 & 29.353 & 0.000052 \\

\midrule

\multirow{5}{*}{Across datasets}
& Accuracy          & 0.977 & 23.459 & 0.000656 \\
& Balanced accuracy & 0.987 & 23.679 & 0.000598 \\
& Macro-F1          & 0.987 & 23.679 & 0.000598 \\
& Macro-recall      & 0.987 & 23.679 & 0.000598 \\
& Macro-precision   & 0.977 & 23.459 & 0.000656 \\

\bottomrule
\end{tabular}
\end{table}

\subsection{Evaluation on NCBI 2026 Collections}
\label{sec:ncbi2026_additional}

To examine CANN beyond the family-filtered benchmarks, we extended the evaluation to two NCBI Virus collections curated through 8 August 2026. The quality-controlled NCBI 2026 collection comprised 14,149 genomes from 272 families. NCBI 2026 All retained the broader collection of 15,098 genomes from 278 families. Their corresponding reference collections were NCBI 2024, containing 12,154 genomes from 199 families, and NCBI 2024 All, containing 13,645 genomes from 209 families.

\newcolumntype{L}[1]{>{\raggedright\arraybackslash}p{#1}}
\begin{table}[!htb]
\centering
\captionsetup{font=small}
\caption{\textbf{Definition of the NCBI 2026 evaluation tests.}
The same definitions were used for NCBI 2026 and NCBI 2026 All with their corresponding 2024 reference collections. No minimum family-size threshold was applied in any test.}
\label{tab:ncbi2026_test_definitions}
\fontsize{10}{12}\selectfont
\setlength{\tabcolsep}{3.5pt}
\renewcommand{\arraystretch}{1.12}
\begin{tabular}{@{}cL{0.51\columnwidth}L{0.27\columnwidth}@{}}
\toprule
Test & Evaluation design & Purpose \\
\midrule
1 &
Only the corresponding 2024 collection is used to train CANN and
select the model. CANN then classifies all genomes present in 2026 but
absent from 2024, including genomes from families not represented in
the 2024 reference. &
Measures performance when some test families are absent from the
reference. \\
\addlinespace[2pt]
2 &
The complete 2026 collection is evaluated using stratified
five-fold cross-validation repeated across 30 random seeds.
Within each fold, training and validation subsets are used for
model fitting and checkpoint selection, and the held-out fold is
used for evaluation. &
Measures performance across the complete 2026 family distribution. \\
\addlinespace[2pt]
3 &
As in Test 1, only the corresponding 2024 collection is used to train CANN and select the model. CANN then classifies only the 2026-only genomes whose families are represented in the 2024 reference. &
Measures performance when the correct family label is available in the
reference. \\
\bottomrule
\end{tabular}
\end{table}

Three complementary evaluations were designed to distinguish the effects of incomplete reference-family coverage from performance across the complete 2026 family distributions. Their data partitions, reference conditions, and objectives are summarized in Table~\ref{tab:ncbi2026_test_definitions}. The corresponding classification results are reported in Table~\ref{tab:ncbi2026_additional}.

\begin{table}[!htbp]
\centering
\captionsetup{font=small}
\caption{\textbf{CANN performance on the NCBI 2026 collections.}
Values are percentages. BA equals macro-recall; F1 and precision are macro-averaged. Standard deviations and encoder-specific results are reported in the Supplementary Table S3.}
\label{tab:ncbi2026_additional}
\fontsize{9.}{11.5}\selectfont
\setlength{\tabcolsep}{3.5pt}
\renewcommand{\arraystretch}{1.05}
\begin{tabular}{@{}llc@{\hspace{8pt}}r@{\hspace{8pt}}r@{\hspace{8pt}}r@{\hspace{8pt}}r@{}}
\toprule
Dataset & Test & Rule & ACC & BA & F1 & Prec. \\
\midrule
\multirow{6}{*}{NCBI 2026}
& Test 1 & 1-NN & 26.20 & 35.70 & 27.07 & 25.38 \\
& Test 1 & 5-NN & 26.12 & 35.68 & 26.86 & 25.11 \\
\addlinespace[1pt]
& Test 2 & 1-NN & 90.09 & 71.76 & 73.89 & 80.19 \\
& Test 2 & 5-NN & 89.22 & 60.71 & 62.44 & 69.19 \\
\addlinespace[1pt]
& \textbf{Test 3} & \textbf{1-NN} & \textbf{88.82} & \textbf{80.69} & \textbf{79.00} & \textbf{81.16} \\
& \textbf{Test 3} & \textbf{5-NN} & \textbf{88.56} & \textbf{80.63} & \textbf{78.95} & \textbf{81.14} \\
\midrule
\multirow{6}{*}{NCBI 2026 All}
& Test 1 & 1-NN & 38.16 & 39.29 & 33.12 & 32.34 \\
& Test 1 & 5-NN & 38.06 & 39.37 & 32.94 & 32.16 \\
\addlinespace[1pt]
& Test 2 & 1-NN & 90.27 & 71.89 & 73.99 & 80.17 \\
& Test 2 & 5-NN & 89.48 & 60.82 & 62.54 & 69.27 \\
\addlinespace[1pt]
& \textbf{Test 3} & \textbf{1-NN} & \textbf{89.12} & \textbf{81.44} & \textbf{80.33} & \textbf{82.86} \\
& \textbf{Test 3} & \textbf{5-NN} & \textbf{88.87} & \textbf{81.61} & \textbf{80.51} & \textbf{83.21} \\
\bottomrule
\end{tabular}
\end{table}

For NCBI 2026 All, Test 1 comprised 1,567 genomes from 114 families. Of these, 896 genuinely new-family genomes belonged to 59 genuinely new families absent from the NCBI 2024 All reference. The remaining 671 known-family genomes belonged to 55 known families. CANN achieved 38.16\% accuracy with 1-NN and 38.06\% with 5-NN. Because the classifier could not predict family labels absent from the reference, the maximum possible accuracy was $671/1567=42.82\%$, even if every supported-family genome was classified correctly. The low score must therefore be interpreted in the context of incomplete reference-family coverage.

Test 2 evaluated all 15,098 NCBI 2026 All genomes and all 278 families. CANN achieved 90.27\% accuracy with 1-NN and 89.48\% with 5-NN. Increasing the neighbourhood size reduced balanced accuracy from 71.89\% to 60.82\%, macro-F1 from 73.99\% to 62.54\%, and macro-precision from 80.17\% to 69.27\%. Thus, a 0.79-percentage-point decrease in accuracy was accompanied by much larger losses in family-averaged metrics. This difference is consistent with the strong class imbalance in the complete collection. Small families may have few representatives in individual folds, allowing 5-NN predictions to be influenced by more abundant families.

Test 3 removed the 896 new-family genomes from the NCBI 2024 All reference. The resulting evaluation contained 671 validation genomes from 55 represented families. CANN reached 89.12\% accuracy with 1-NN and 88.87\% with 5-NN. Although 5-NN reduced accuracy by 0.25 percentage points, it increased balanced accuracy by 0.17 points, macro-F1 by 0.18 points, and macro-precision by 0.35 points. These results show that CANN retained strong family-discriminative information when the correct family label was available in the reference collection.

The encoder analysis also supported combining the two CANN representations. Across all three NCBI 2026 All tests, the equally weighted CANN distance outperformed both encoder-derived distance sources for every reported metric under both neighbourhood rules. For example, in Test 2 with 5-NN, the Transformer-derived accuracy was 87.47\%, compared with 83.40\% for the CNN-derived distance. The CNN and Transformer produced balanced accuracies of 58.45\% and 57.97\%, respectively, and macro-F1 scores of 59.95\% and 59.62\%. CANN achieved 89.48\% accuracy, 60.82\% balanced accuracy, and 62.54\% macro-F1. Complete encoder-specific results and standard deviations are provided in the Supplementary Section S4 and Supplementary Table S3.

The NCBI 2026 collection showed the same overall pattern. Test 1 contained 2,102 unseen genomes from 113 new families, including 1,482 genuinely new-family genomes from 63 families absent from the NCBI 2024 reference. The remaining 620 known-family genomes from 50 represented families formed Test 3. CANN achieved 26.20\% and 26.12\% accuracy with 1-NN and 5-NN in Test 1, 90.09\% and 89.22\% in Test 2, and 88.82\% and 88.56\% in Test 3. The complete balanced accuracy, macro-F1, macro-recall, and macro-precision results are given in Table~\ref{tab:ncbi2026_additional}.

Together, these evaluations distinguish two sources of difficulty. Missing reference families constrained Test 1, whereas the complete and highly imbalanced collection made 5-NN less reliable for family-averaged metrics in Test 2. When the family label was represented in the reference, CANN maintained high accuracy and strong family-balanced performance.

%\subsection{Family-specific Analysis of Classification Error} To better understand the classification errors observed for the NCBI 2026 data, we performed a family-specific analysis using the Test 3 setting, in which the frozen NCBI 2024 All dataset was used as the reference set and the 2026 Validation set was used for testing. The 2026 Validation set contains genomes that were absent from NCBI 2024 All but belong to families already represented in the 2024 reference set. Genuinely new families were therefore excluded from this analysis.

%A query genome is taken from NCBI 2026 All. Its true family in the 2026 dataset is denoted as family A, while the incorrect 2024 All reference family to which the query genome is misclassified is denoted as family B.

\begin{figure}[htb!]
    \centering
    \captionsetup{font=small}
    \includegraphics[width=\textwidth]{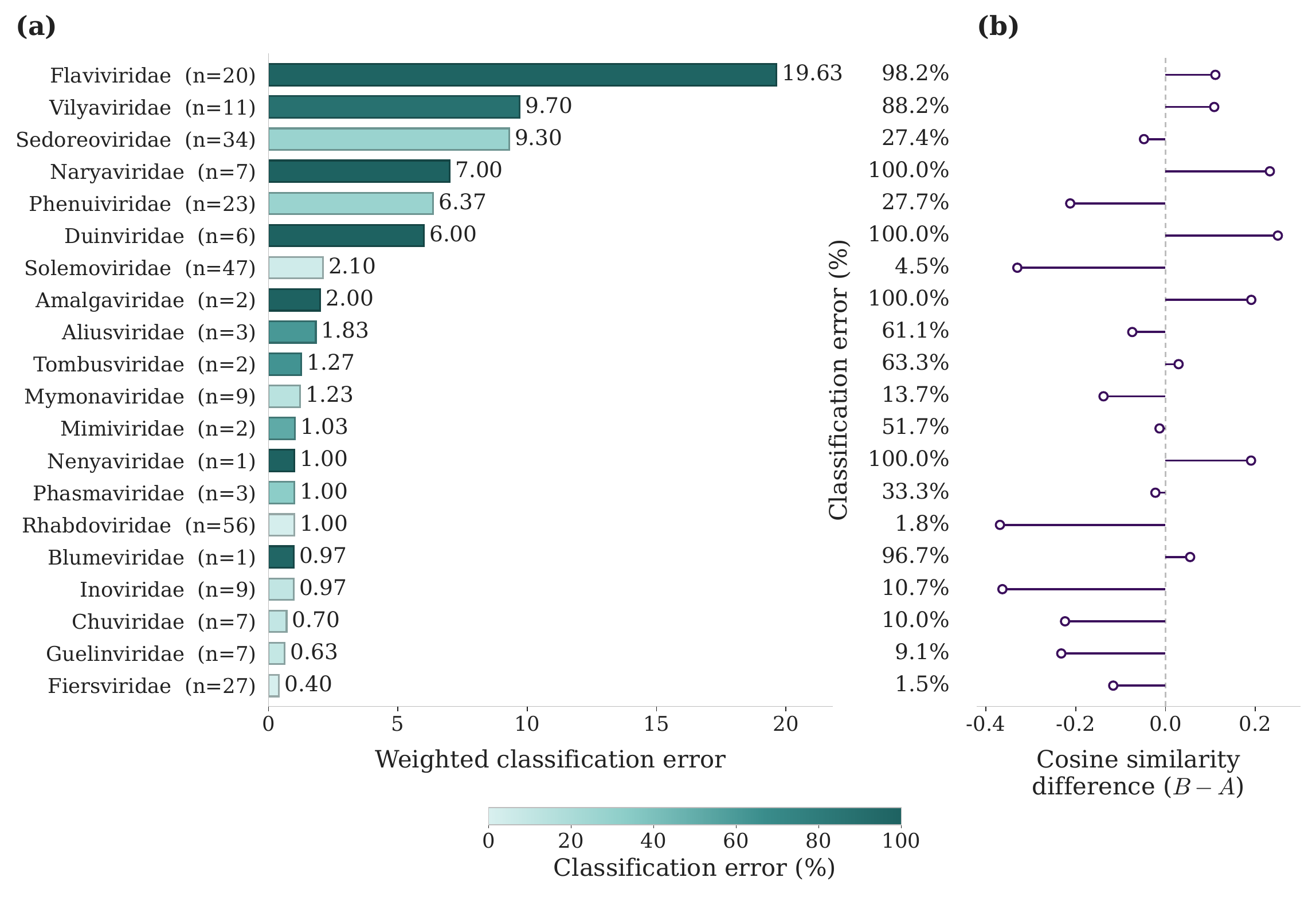}
    \caption{\textbf{Family-specific errors and cosine similarity differences for the 20 families with the largest weighted errors.}
    Bars show the weighted error, defined as the average number of misclassified genomes per random seed. Bar shading represents the family-specific classification error, and $n$ gives the number of Test 3 genomes in each family. The line shows
    $\Delta S=S_B-S_A$, where $A$ is the true family and $B$ is the most frequently predicted incorrect family. Positive values indicate greater mean similarity to $B$, whereas negative values indicate greater mean similarity to $A$. Results are based on CANN 5-NN predictions across 30 random seeds.}
    \label{fig:familyerror}
\end{figure}

\subsubsection{Family-Specific Classification Errors.}
\label{sec:family_error}

Aggregate scores do not show whether classification errors are concentrated in particular viral families. We therefore examined the CANN 5-NN predictions for the 671 NCBI 2026 All Test 3 genomes, representing 55 families. Predictions were collected across 30 random seeds. For each seed, the CNN- and Transformer-derived cosine distances were averaged with equal weights before classification.

For each family, the classification error is the percentage of incorrect predictions across its genomes and all 30 seeds. The weighted error is the average number of incorrectly classified genomes from that family per seed. It accounts for both the error rate and the number of genomes. Family $A$ denotes the true family, whereas family $B$ denotes its most frequently predicted incorrect family. Predicted as $B$ is the percentage of all predictions for family $A$ that are assigned to family $B$.

The similarity difference, denoted by $\Delta S$, compares the mean similarity of the query genomes to families $B$ and $A$. A positive value indicates greater similarity to the confused family $B$, whereas a negative value indicates greater similarity to the correct family $A$. This comparison is used only to interpret the classification errors and does not determine the 5-NN prediction. Complete mathematical definitions are provided in Supplementary Section S3.

Figure~\ref{fig:familyerror} presents the 20 families with the largest weighted errors, and Table~\ref{tab:family_error} provides their numerical results. Flaviviridae produced the largest weighted error. Its 20 genomes had an error of 98.17\%, corresponding to an average of 19.63 misclassified genomes per seed. Chrysoviridae was its most frequent incorrect prediction, and the positive similarity difference of $0.1111$ indicates that Chrysoviridae was closer on average as demonstrated in Table~\ref{tab:family_error}. However, only 39.67\% of all Flaviviridae predictions were assigned to Chrysoviridae. Its remaining errors were therefore distributed among several families.

Vilyaviridae showed a more concentrated confusion pattern. Its 11 genomes had an error of 88.18\% and a weighted error of 9.70. Circoviridae accounted for 79.39\% of all predictions, while the similarity difference was positive at $0.1086$. Thus, both the prediction frequency and the similarity comparison indicate a consistent confusion with Circoviridae.

A positive similarity difference was not observed for every high-error family. Sedoreoviridae contained 34 genomes and had an error of 27.35\%, resulting in the third-largest weighted error of 9.30. Nevertheless, its similarity difference relative to Spinareoviridae was slightly negative at $-0.0479$. Phenuiviridae showed a similar pattern, with an error of 27.68\%, a weighted error of 6.37, and a similarity difference of $-0.2122$ relative to Fimoviridae. The query genomes from these families remained closer to their correct families on average. Their errors are therefore more consistent with the local composition of individual 5-NN neighborhoods than with an overall shift toward the confused family.

Family size also changes how percentage errors should be interpreted. Naryaviridae and Duinviridae both had 100\% error, but they contained only seven and six genomes, giving weighted errors of 7.00 and 6.00, respectively. In contrast, Rhabdoviridae contained 56 genomes but had an error of only 1.79\% and a weighted error of 1.00. Its strongly negative similarity difference of $-0.3687$ was also consistent with clear separation from its main confused family. These examples show that neither percentage error nor family size alone provides a complete description of family-level performance.

Together, the family-specific error, weighted error, confusion frequency, and similarity difference distinguish two forms of classification difficulty. Positive similarity differences identify families whose genomes are closer to a competing family on average. Negative differences identify cases in which the correct family remains closer overall, although some genomes acquire incorrect local neighborhoods during 5-NN classification.

\begin{table*}[!htb]
\centering
\captionsetup{font=small}
\caption{\textbf{Family-specific classification statistics for the 20 families with the largest weighted errors.}
Results are based on CANN 5-NN predictions for the NCBI 2026 All
Test 3 genomes across 30 random seeds. Family $A$ is the true family,
and family $B$ is the most frequently predicted incorrect family.
Weighted error is the average number of misclassified genomes per seed.
Predicted as $B$ is the percentage of all predictions for family $A$
assigned to family $B$. The similarity difference is defined as
$\Delta S=\mathrm{Cosine}(B)-\mathrm{Cosine}(A)$. Positive values
indicate greater similarity to $B$, whereas negative values indicate
greater similarity to $A$.}
\label{tab:family_error}
\fontsize{9.7}{12}\selectfont
\setlength{\tabcolsep}{3pt}
\renewcommand{\arraystretch}{1.05}
\begin{tabular}{@{}lccclcc@{}}
\toprule
\textbf{Family $A$} &
\textbf{$n$} &
\shortstack{\textbf{Error}\\\textbf{(\%)}} &
\shortstack{\textbf{Weighted}\\\textbf{error}} &
\textbf{Family $B$} &
\shortstack{\textbf{Predicted as}\\\textbf{$B$ (\%)}} &
\shortstack{\boldmath$\Delta S$\unboldmath\\\textbf{($B-A$)}} \\
\midrule
Flaviviridae   & 20 & 98.17  & 19.63 & Chrysoviridae       & 39.67 &  0.1111 \\
Vilyaviridae   & 11 & 88.18  & 9.70  & Circoviridae        & 79.39 &  0.1086 \\
Sedoreoviridae & 34 & 27.35  & 9.30  & Spinareoviridae     & 22.16 & -0.0479 \\
Naryaviridae   & 7  & 100.00 & 7.00  & Circoviridae        & 25.24 &  0.2325 \\
Phenuiviridae  & 23 & 27.68  & 6.37  & Fimoviridae         & 10.87 & -0.2122 \\
Duinviridae    & 6  & 100.00 & 6.00  & Fiersviridae        & 86.67 &  0.2502 \\
Solemoviridae  & 47 & 4.47   & 2.10  & Fiersviridae        & 1.84  & -0.3300 \\
Amalgaviridae  & 2  & 100.00 & 2.00  & Botourmiaviridae    & 51.67 &  0.1911 \\
Aliusviridae   & 3  & 61.11  & 1.83  & Chuviridae          & 22.22 & -0.0740 \\
Tombusviridae  & 2  & 63.33  & 1.27  & Totiviridae         & 58.33 &  0.0292 \\
Mymonaviridae  & 9  & 13.70  & 1.23  & Rhabdoviridae       & 7.04  & -0.1380 \\
Mimiviridae    & 2  & 51.67  & 1.03  & Baculoviridae       & 28.33 & -0.0133 \\
Nenyaviridae   & 1  & 100.00 & 1.00  & Circoviridae        & 60.00 &  0.1906 \\
Phasmaviridae  & 3  & 33.33  & 1.00  & Peribunyaviridae    & 31.11 & -0.0225 \\
Rhabdoviridae  & 56 & 1.79   & 1.00  & Mymonaviridae       & 1.25  & -0.3687 \\
Blumeviridae   & 1  & 96.67  & 0.97  & Fiersviridae        & 63.33 &  0.0549 \\
Inoviridae     & 9  & 10.74  & 0.97  & Polydnaviriformidae & 10.00 & -0.3632 \\
Chuviridae     & 7  & 10.00  & 0.70  & Lispiviridae        & 6.19  & -0.2237 \\
Guelinviridae  & 7  & 9.05   & 0.63  & Rountreeviridae     & 4.76  & -0.2321 \\
Fiersviridae   & 27 & 1.48   & 0.40  & Steitzviridae       & 1.36  & -0.1164 \\
\bottomrule
\end{tabular}
\end{table*}

\subsubsection{Feature-Block Ablation.}
\label{sec:feature_ablation}

To examine which feature scales CANN relies on, we also performed a feature-block ablation analysis using 1-NN classification in the Test 3 setting for both NCBI 2026 collections. Each facet block at $k=3,4,5$ and each $k$-mer block at $k=6,7$ was masked separately. Masking replaces the block with its reference-set mean in both the query and reference genomes. The remaining blocks and encoder parameters are unchanged. The ablation effect measures the decrease from the full CANN score, expressed in percentage points. Positive values indicate lower performance after masking, whereas negative values indicate an improvement.

\begin{figure}[htb!]
\centering
\captionsetup{font=small}
\includegraphics[width=0.95\linewidth]{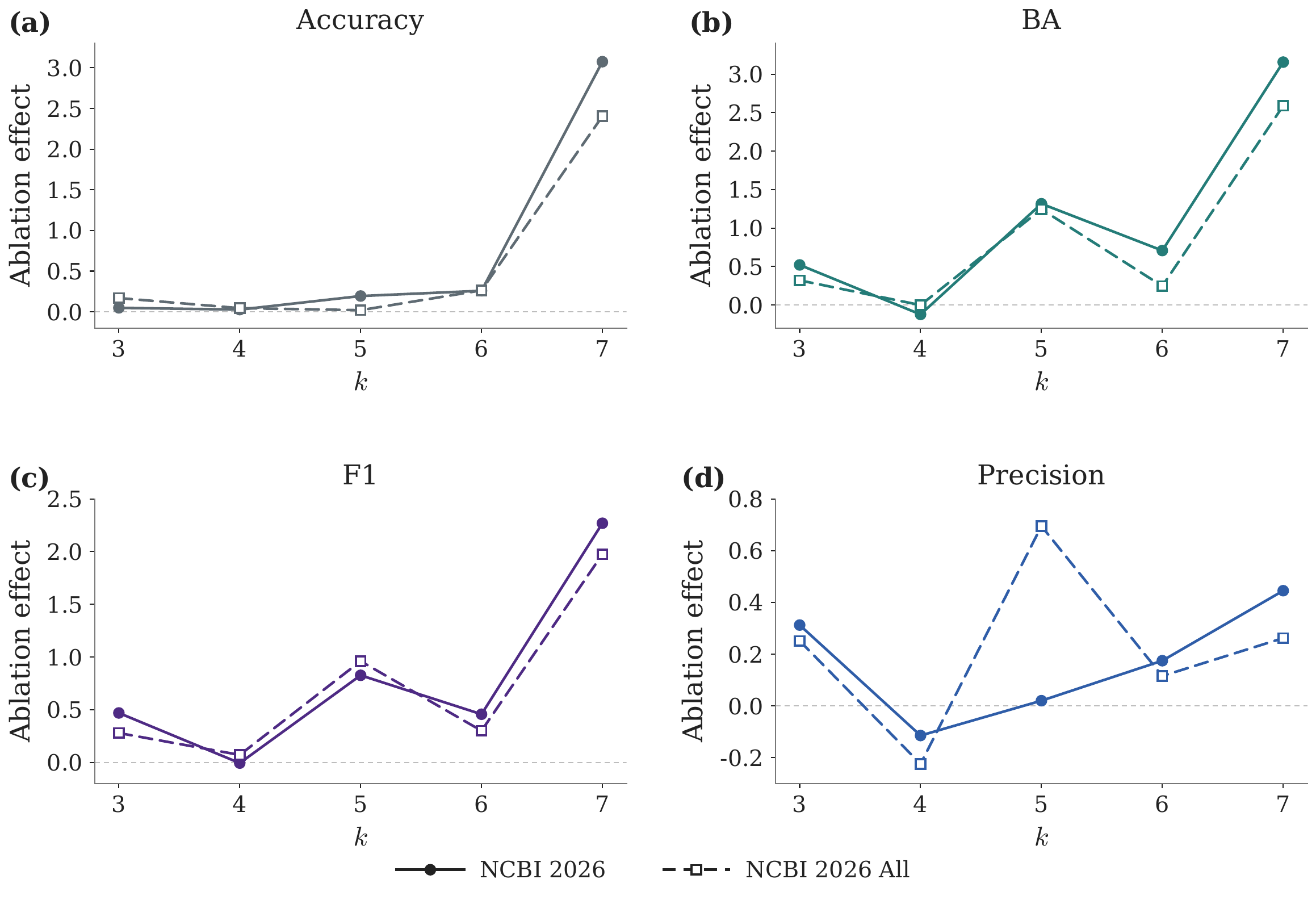}
\caption{\textbf{Sensitivity of CANN to individual feature blocks.}
One facet block at $k=3,4,5$ or one $k$-mer block at $k=6,7$
is masked at a time by replacing it with its reference-set mean
in both query and reference genomes. The encoders are not retrained.
Results use CANN 1-NN classification in the Test 3 setting.
The ablation cohorts contain 619 genomes from 49 families for
NCBI 2026 and 671 genomes from 55 families for NCBI 2026 All.
Ablation effects are calculated within each seed as the full-model score minus the masked-model score, expressed in percentage points and averaged across 30 seeds. Positive values indicate reduced performance after masking; negative values indicate improvement. Metrics are accuracy, balanced accuracy (BA), macro-F1, and macro-precision.}
\label{fig:ablationtests}
\end{figure}

Figure~\ref{fig:ablationtests} shows that masking the $k$-mer block at $k=7$ produces the largest mean decrease in accuracy, balanced accuracy, and macro-F1 on both datasets. Accuracy decreases by 3.07 percentage points on NCBI 2026 and 2.40 percentage points on NCBI 2026 All. Among the facet blocks, masking $k=5$ has the largest effect on balanced accuracy and macro-F1, despite its smaller effect on overall accuracy. This distinction suggests that the facet information at this scale is particularly relevant to average family-level recognition.

The other blocks produce smaller and more metric-dependent changes. Some effects are close to zero or negative, particularly for facet $k=4$. Thus, masking a feature block does not uniformly reduce every performance measure. Overall, the results indicate the strongest reliance on the $k=7$ composition features, with facet $k=5$ also supporting family-averaged performance. Because the encoders are not retrained, these findings describe the sensitivity of the trained CANN model rather than identify an optimal feature subset.

\section{Methods and Materials}

\subsection{Persistent Stanley-Reisner Facet Representation}
\label{subsec:cakr_cann}

CANN uses the facet-based component of the commutative algebra \(k\)-mer representation (CAKR) framework to encode the positional organization of recurring nucleotide patterns \cite{suwayyid2025persistent,suwayyid2026cakr}. The construction employed here consists of four steps: extracting the occurrence positions of each \(k\)-mer, constructing a Vietoris-Rips filtration on those positions, associating Stanley-Reisner facet data with the filtered complexes, and assembling the resulting facet-number values into a fixed numerical representation.

\subsubsection{Occurrence-position representation of k-mers}

Let $\mathcal{A}=\{\mathrm{A},\mathrm{C},\mathrm{G},\mathrm{T}\}$ be the nucleotide alphabet, and let $S=s_1s_2\cdots s_N\in\mathcal{A}^N$ be a genome sequence. For a fixed integer \(k>0\), a \(k\)-mer is a word $\mathbf{x}=x_1x_2\cdots x_k\in\mathcal{A}^k$. For each \(\mathbf{x}\), define the indicator function
\[
\delta_{\mathbf{x}}(\mathbf{y})
=
\begin{cases}
1, & \mathbf{y}=\mathbf{x},\\
0, & \mathbf{y}\neq\mathbf{x},
\end{cases}
\qquad
\mathbf{y}\in\mathcal{A}^k.
\]
The set of starting positions at which \(\mathbf{x}\) occurs in \(S\) is
\begin{equation}
\label{eq:cann-kmer-position-set}
S^{\mathbf{x}}
=
\left\{
i\in\{1,\ldots,N-k+1\}
\;\middle|\;
\delta_{\mathbf{x}}
\bigl(s_i s_{i+1}\cdots s_{i+k-1}\bigr)=1
\right\}.
\end{equation}
Thus, $S^{\mathbf{x}}=\{p_1,\ldots,p_m\}$ retains the positions \(p_1,\ldots,p_m\), rather than only the occurrence count \(m\). This approach allows two sequences having the same number of occurrences of a given \(k\)-mer to be distinguished when those occurrences are arranged differently along the genome.

The occurrence positions are regarded as points on the real line, with pairwise distance
\begin{equation}
\label{eq:cann-kmer-distance}
d^{\mathbf{x}}(p_a,p_b)
=
|p_a-p_b|.
\end{equation}
Equivalently, the corresponding distance matrix is $\displaystyle D^{\mathbf{x}}=\left(|p_a-p_b|\right)_{1\leq a,b\leq m}$. These one-dimensional point clouds form the input to the persistent Stanley-Reisner construction.

\subsubsection{Vietoris-Rips filtration and Stanley-Reisner facets}

For a filtration value \(r\geq 0\), let
\(\Delta_{\mathbf{x}}^{r}\) denote the Vietoris-Rips complex built from the occurrence positions \(S^{\mathbf{x}}\). Explicitly,
\begin{equation}
\label{eq:cann-rips-complex}
\Delta_{\mathbf{x}}^{r}
=
\left\{
\sigma\subseteq S^{\mathbf{x}}
\;\middle|\;
|p-q|\leq r
\text{ for every }p,q\in\sigma
\right\}.
\end{equation}
Hence a finite collection of occurrence positions forms a face when all of its members are mutually within the filtration threshold \(r\). If \(r\leq r'\), then $\Delta_{\mathbf{x}}^{r}\subseteq\Delta_{\mathbf{x}}^{r'}$, so the family $\left(\Delta_{\mathbf{x}}^{r}\right)_{r\geq 0}$ forms an increasing filtration. A facet of \(\Delta_{\mathbf{x}}^{r}\) is an inclusion-maximal face. Denote the set of facets by $\mathcal{F}\!\left(\Delta_{\mathbf{x}}^{r}\right)$. To connect this combinatorial structure with commutative algebra, write the vertices of \(\Delta_{\mathbf{x}}^{r}\) as \(v_1,\ldots,v_m\), let $R_{\mathbf{x}}=\Bbbk[z_1,\ldots,z_m]$ be the standard graded polynomial ring over a field \(\Bbbk\), and define the Stanley-Reisner ideal
\begin{equation}
\label{eq:cann-sr-ideal}
I_{\mathbf{x}}^{r}
=
\left\langle
z_{i_1}\cdots z_{i_q}
\;\middle|\;
\{v_{i_1},\ldots,v_{i_q}\}
\notin
\Delta_{\mathbf{x}}^{r}
\right\rangle.
\end{equation}
Its Stanley--Reisner ring is $\Bbbk[\Delta_{\mathbf{x}}^{r}]=R_{\mathbf{x}}/I_{\mathbf{x}}^{r}$. The Stanley-Reisner ideal has the canonical irredundant decomposition
\begin{equation}
\label{eq:cann-facet-decomposition}
I_{\mathbf{x}}^{r}
=
\bigcap_{\sigma\in
\mathcal{F}(\Delta_{\mathbf{x}}^{r})}
P_{\sigma},
\qquad
P_{\sigma}
=
\left(
z_j
\;\middle|\;
v_j\notin\sigma
\right).
\end{equation}
Thus, facets of the filtered simplicial complex index the prime components of its Stanley-Reisner ideal. For each dimension \(i\geq 0\), define
\begin{equation}
\label{eq:cann-dim-facet-primes}
\mathcal{P}_{i,\mathbf{x}}^{r}
=
\left\{
P_{\sigma}
\;\middle|\;
\sigma\in
\mathcal{F}(\Delta_{\mathbf{x}}^{r}),
\ \dim(\sigma)=i
\right\}.
\end{equation}
For \(r\leq r'\), the persistent facet-prime set may be written as
\begin{equation}
\label{eq:cann-persistent-facet-primes}
\mathcal{P}_{i,\mathbf{x}}^{r,r'}
=
\mathcal{P}_{i,\mathbf{x}}^{r}
\cap
\mathcal{P}_{i,\mathbf{x}}^{r'},
\end{equation}
with corresponding persistent facet number
\begin{equation}
\label{eq:cann-persistent-facet-number}
\mathcal{F}_{i,\mathbf{x}}^{r,r'}
=
\left|
\mathcal{P}_{i,\mathbf{x}}^{r,r'}
\right|.
\end{equation}
CANN uses the corresponding facet-number curve, that is, the number of active facet components evaluated at selected filtration locations. We write
\begin{equation}
\label{eq:cann-facet-curve}
\mathcal{F}_{i,\mathbf{x}}(r)
=
\mathcal{F}_{i,\mathbf{x}}^{r,r}
=
\left|
\mathcal{P}_{i,\mathbf{x}}^{r}
\right|
\end{equation}
for the diagonal facet number at scale \(r\).

\subsubsection{Facet-number features used by CANN}

CANN retains the dimension-\(0\) facet component. Accordingly, for a fixed \(k\)-mer \(\mathbf{x}\in\mathcal{A}^k\), its CANN algebraic feature is obtained by sampling \(\mathcal{F}_{0,\mathbf{x}}(r)\) at a selected set of filtration values \(r_{k,1},r_{k,2},\ldots,r_{k,m_k}\) in increasing order.
Thus, the feature associated with \(\mathbf{x}\) is
\begin{equation}
	\label{eq:cann-single-kmer-facet-feature}
	\boldsymbol{\phi}_{\mathbf{x}}^{(k)}(S)
	=
	\left(
	\mathcal{F}_{0,\mathbf{x}}(r_{k,1}),
	\mathcal{F}_{0,\mathbf{x}}(r_{k,2}),
	\ldots,
	\mathcal{F}_{0,\mathbf{x}}(r_{k,m_k})
	\right)
	\in\mathbb{R}^{m_k}.
\end{equation}

A fixed ordering of all \(4^k\) nucleotide \(k\)-mers is used. Concatenating the \(m_k\)-dimensional facet feature of every \(\mathbf{x}\in\mathcal{A}^k\) produces the CANN facet block
\begin{equation}
	\label{eq:cann-facet-block}
	\Phi_k(S)
	=
	\left(
	\boldsymbol{\phi}_{\mathbf{x}}^{(k)}(S)
	\right)_{\mathbf{x}\in\mathcal{A}^k}
	\in
	\mathbb{R}^{4^k\times m_k}.
\end{equation}
If a \(k\)-mer does not occur in \(S\), its corresponding row is set to zero.

CANN computes these blocks for \(k=3\), \(k=4\), and \(k=5\). Consequently, the three inputs have dimensions
\[
\Phi_3(S)\in\mathbb{R}^{64\times m_3},
\qquad
\Phi_4(S)\in\mathbb{R}^{256\times m_4},
\qquad
\Phi_5(S)\in\mathbb{R}^{1024\times m_5}.
\]
Here, $m_k$ denotes the number of sampled filtration values.
In our implementation, the filtration values are $0$ and $4^k$,
giving $m_3=m_4=m_5=2$. The three facet blocks therefore contain
128, 512, and 2,048 scalar features, respectively.
These blocks encode the facet behavior of the occurrence-position sets separately for every nucleotide word at each selected \(k\)-mer length and across multiple stages of the filtration. In this way, each row of \(\Phi_k(S)\) describes not only the facet information associated with a particular \(k\)-mer, but also its evolution as the filtration parameter varies over the selected filtration locations.

Because the raw facet-number values may vary in scale across \(k\)-mers and filtration locations, CANN applies elementwise min-max normalization to each facet block. To preserve the separation between training and held-out data, the normalization parameters are determined solely from the training genomes of the corresponding cross-validation fold.

For a fixed \(k\), let
\[
m_{k,\mathbf{x},\ell}
=
\min_{S\in\mathcal{T}}
\Phi_k(S)_{\mathbf{x},\ell},
\qquad
M_{k,\mathbf{x},\ell}
=
\max_{S\in\mathcal{T}}
\Phi_k(S)_{\mathbf{x},\ell},
\]
where \(\mathcal{T}\) denotes the training subset and \(\ell\in\{1,\ldots,m_k\}\) indexes the selected filtration values. The normalized feature is defined by
\begin{equation}
	\label{eq:cann-facet-normalization}
	\widetilde{\Phi}_k(S)_{\mathbf{x},\ell}
	=
	\frac{
		\Phi_k(S)_{\mathbf{x},\ell}
		-
		m_{k,\mathbf{x},\ell}
	}{
		M_{k,\mathbf{x},\ell}
		-
		m_{k,\mathbf{x},\ell}
	},
	\qquad
	\ell=1,\ldots,m_k.
\end{equation}

When \(M_{k,\mathbf{x},\ell}=m_{k,\mathbf{x},\ell}\), the corresponding normalized coordinate is taken to be zero. The normalized values lie in the interval \([0,1]\). The same training-derived minima and maxima are used for the validation, reference, and held-out test genomes. In this way, the normalization is determined entirely by the training data and does not incorporate information from the held-out genomes.

The three normalized matrices \(\widetilde{\Phi}_3(S)\), \(\widetilde{\Phi}_4(S)\), and \(\widetilde{\Phi}_5(S)\) constitute the persistent commutative-algebraic component of the CANN input representation. They are subsequently combined with within-genome normalized conventional \(k\)-mer frequency blocks for \(k=6\) and \(k=7\).

\subsubsection{Illustrative Example}
\label{subsubsec:illustrative_example}

Consider three hypothetical genomes of equal length: reference genomes $S_A$ and $S_B$ from families A and B, respectively, and an unlabelled test genome $Q$. A $k$-mer is a nucleotide word of length $k$. Suppose that the $3$-mer $\mathbf{x}=\texttt{ATA}$ begins at positions $S^{\mathbf{x}}_{A}=\{2,4,6\}$, $S^{\mathbf{x}}_{B}=\{2,20,38\}$, and $Q^{\mathbf{x}}=\{3,5,7\}$. Its occurrence count is three in every genome, giving identical
relative frequencies because the genomes have equal lengths. The frequency of this word alone therefore cannot distinguish the genomes. Nevertheless, its occurrences are closely spaced in $S_A$ and $Q$ and widely separated in $S_B$.

To describe this organization, each occurrence position is treated as a vertex. For illustration, two positions $p_i$ and $p_j$ are connected at distance threshold $r$ when $|p_i-p_j|\leq r$. A collection forms a face when every pair of its vertices satisfies this criterion. A facet is an inclusion-maximal face: no additional occurrence position can be added while preserving the pairwise distance criterion. Facets can overlap, and an isolated vertex also constitutes a facet.

For $S_A$, the pairwise separations are $2$, $2$, and $4$. At $r=3$, the pairs $\{2,4\}$ and $\{4,6\}$ are facets, whereas $\{2,4,6\}$ is not a face because $|2-6|=4>3$. The singleton $\{4\}$ is a face but not a facet because it is contained in larger faces. At $r=5$, all three positions are
mutually connected, making $\{2,4,6\}$ the sole facet; the earlier pairs remain faces but cease to be maximal. The test genome $Q$ has the same pairwise separations and follows the same construction, with every position shifted by one. In contrast, the separations in $S_B$ are $18$, $18$, and $36$, so its three occurrences remain isolated at both thresholds (Table~\ref{tab:illustrative_facets}).

\begin{table}[htbp]
\centering
\captionsetup{font=small}
\small
\renewcommand{\arraystretch}{1.15}
\setlength{\tabcolsep}{5.5pt}
\caption{Facets of the \texttt{ATA} occurrence positions at two
illustrative distance thresholds. Each set in braces represents
one facet.}
\label{tab:illustrative_facets}
\begin{tabular}{lcc}
\toprule
Genome & Facets at $r=3$ & Facets at $r=5$ \\
\midrule
$S_A$ & $\{2,4\},\,\{4,6\}$ & $\{2,4,6\}$ \\
$S_B$ & $\{2\},\,\{20\},\,\{38\}$
      & $\{2\},\,\{20\},\,\{38\}$ \\
$Q$   & $\{3,5\},\,\{5,7\}$ & $\{3,5,7\}$ \\
\bottomrule
\end{tabular}
\end{table}

Repeating this construction across increasing thresholds describes how the organization of occurrences changes with distance scale. Facets index the prime components of the corresponding Stanley-Reisner ideal, connecting this construction to commutative algebra~\cite{suwayyid2026cakr}. Numerical descriptors derived from these facets summarize positional information that complements word abundance. In this example, $Q$ and $S_A$ have identical relative spacing, whereas the occurrences in $S_B$ form larger facets only at substantially greater thresholds.

In CANN, facet descriptors for $k=3,4,5$ are concatenated with $k$-mer frequency features for $k=6,7$. The combined feature vector is supplied independently to CNN and Transformer encoders. The CNN learns local combinations of neighbouring entries within the feature array, while the Transformer's self-attention mechanism learns relationships among feature tokens across the representation. Using known family labels during training, each encoder maps a genome to a compact numerical vector, or embedding. Test and reference genomes are subsequently compared within each encoder's embedding space using cosine distance, with smaller distances indicating greater similarity.

For illustration, suppose that the trained CNN and Transformer
encoders, operating on the complete feature vectors, produce
cosine distances from $Q$ to $(S_A,S_B)$ of $(0.08,0.55)$ and
$(0.14,0.40)$, respectively. CANN averages the corresponding
distances with equal weights, giving $(0.08+0.14)/2=0.11$
for $S_A$ and $(0.55+0.40)/2=0.475$ for $S_B$. Because $0.11<0.475$, 1-nearest-neighbour classification assigns $Q$ to family A in this two-reference example. For a larger reference collection, 5-nearest-neighbour classification uses the same consensus distances to identify the five closest reference genomes and selects the family receiving the most votes. For example, neighbour labels $(A,A,B,A,B)$ yield family A with three votes. Thus, sequence abundance and multiscale positional information are transformed into learned representations, whose combined distances support alignment-free assignment to a family represented in the labelled reference collection.

\section{Deep Learning Modeling}
\label{sec:deep_learning_modeling}

CANN trained independent CNN and Transformer encoders on persistent facet descriptors at $k=3,4,5$ and $k$-mer frequencies at $k=6,7$. Facet descriptors were scaled using training-derived minima and maxima, with values clipped to $[0,1]$. The same scaling was applied to validation and test genomes. Each $k$-mer count vector was normalized by its sum.

\begin{table}[!htb]
\centering
\captionsetup{font=small}
\caption{\textbf{CANN architecture and training settings.}
Shared settings apply to both encoders. CNN channels, kernel widths,
and strides are listed along the main convolutional path.
The middle two convolutions form the residual block.
Training limits and learning-rate scheduling apply during model
selection.}
\label{tab:cann-model-parameters}
\small
\setlength{\tabcolsep}{4pt}
\renewcommand{\arraystretch}{1.06}
\begin{tabular}{
@{}>{\raggedright\arraybackslash}p{0.30\linewidth}
>{\raggedright\arraybackslash}p{0.64\linewidth}@{}}
\toprule
\textbf{Component} & \textbf{Setting} \\
\midrule
Genome embedding
& 128 dimensions \\
CNN convolutions
& Channels: $16,32,32,32$;
  kernels: $7,5,5,5$;
  strides: $4,1,1,2$;
  batch normalization and ReLU \\
CNN pooling
& Max pooling with width 2 and stride 2;
  adaptive average pooling to 4 output positions \\
Transformer
& 2 layers; 4 attention heads;
  feed-forward dimension 512;
  GELU, residual connections, and layer normalization \\
Projection head
& $128 \rightarrow 128 \rightarrow 128$;
  hidden ReLU activation \\
Dropout
& CNN branches and Transformer input/layers: 0.10;
  classifier: 0.30 \\
\midrule
Optimizer
& AdamW; initial learning rate $5\times10^{-4}$;
  weight decay $10^{-4}$ \\
Contrastive loss
& Weight: 0.10; temperature: 0.07 \\
Batch composition
& Up to 16 families, with 4 genomes per family \\
Training limit
& Maximum 60 epochs;
  early-stopping patience: 10 epochs \\
Learning-rate scheduling
& Reduction factor: 0.5;
  patience: 3 epochs;
  minimum learning rate: $10^{-6}$ \\
Gradient clipping
& Maximum gradient norm: 5 \\
\bottomrule
\end{tabular}
\end{table}

The CNN processed each feature block through a residual convolutional branch, using the two filtration positions as channels for facet inputs. Pooled branch outputs were concatenated and mapped to a genome embedding. The Transformer projected each complete block into a token. Learned positional embeddings identified the blocks, and self-attention modeled relationships among them. Mean pooling produced the Transformer embedding. Both encoder embeddings were batch-normalized. Table~\ref{tab:cann-model-parameters} summarizes the architecture and training settings.

Training combined classification cross-entropy with supervised contrastive loss. A linear classification head predicted family labels, while a separate projection head produced unit-length vectors for contrastive learning. Batches sampled families uniformly and contained equal numbers of genomes per family, with replacement when needed. Both heads were used only during training.

Each encoder's checkpoint is selected using validation accuracy from cosine 1-NN classification, with macro-F1 used to break ties. During checkpoint selection, validation genomes are classified against the training genomes. After selection, the checkpoint is fixed, and the training and validation genomes jointly form the reference collection for held-out testing. The CNN and Transformer encoders produce separate cosine-distance matrices, which are averaged with equal weights before nearest-neighbor classification. Predictions use either 1-NN or unweighted voting among the five nearest references.

For Tests~1 and~3 of the 2026 evaluations, an internal split of the corresponding 2024 reference collection selects the training duration and learning-rate schedule. Each encoder is then initialized independently and trained on the complete 2024 reference. Facet normalization is fitted to that reference, and the resulting encoders, normalization parameters, and reference embeddings remain fixed when classifying the 2026 genomes.

Performance is summarized using accuracy, balanced accuracy,
macro-F1, macro-recall, and macro-precision. Accuracy is the
proportion of correctly classified genomes. For each family,
precision measures the proportion of assigned genomes that belong
to that family, recall measures the proportion of that family's
genomes that are correctly identified, and F1 is the harmonic mean of precision and recall. Macro-averaged metrics give equal weight to every family present in the true test labels, including families absent from the reference. Predicted labels outside this family set are not added as additional classes to the macro-averages, but they still count as incorrect predictions and as false negatives for the true family. Ratios with zero denominators are assigned zero. Balanced accuracy is identical to macro-recall in these evaluations.

\section{Conclusion}
\label{sec:conclusion}

CANN advances alignment-free viral family classification by integrating persistent commutative-algebraic descriptors with neural representation learning. It extends the use of commutative algebra in sequence analysis from distance construction, as in CAKR~\cite{suwayyid2026cakr}, to descriptor construction coupled with neural representation learning. The framework combines multiscale positional information from persistent facet descriptors with complementary $k$-mer composition features, which CNN and Transformer encoders transform into learned genome representations. Averaging cosine distances from the two embedding spaces consistently improved classification over either encoder alone across the four reference benchmarks, with gains in both overall accuracy and macro-averaged performance. Under 5-NN classification, CANN also achieved higher accuracy than the benchmark values available in the literature for CAKR~\cite{suwayyid2026cakr} and all five other alignment-free methods considered.

These findings support the effectiveness of combining algebraic sequence descriptions, supervised learning, and consensus genome comparison within a unified framework. The NCBI 2026 evaluations further supported generalization to additional genomes from families represented in the 2024 reference collection. However, nearest-neighbour assignment remains restricted to families present in the reference. Moreover, the unfiltered evaluation showed that similar overall accuracies for 1-NN and 5-NN can conceal substantial differences in performance across families, emphasizing the importance of evaluating both measures.
%Further development should expand reference coverage, incorporate mechanisms for flagging unsupported assignments, and assess generalization using sequence-clustered and prospective benchmarks.
Together, these results establish CANN as an effective approach to viral family classification and support the broader investigation of persistent algebraic descriptors for learning from genomic sequence organization.

\section*{Data and Code Availability}
The genomic data supporting the findings of this study were
obtained from the National Center for Biotechnology Information
(NCBI). The corresponding sequence records
can be accessed and downloaded through 
\url{https://www.ncbi.nlm.nih.gov/labs/virus/}.
The accession versions and viral family labels used for the
benchmark datasets and all codes needed to evaluate the conclusions in this study are available at \href{https://github.com/mzia-s/CANN}{https://github.com/mzia-s/CANN}.

\section*{Supporting Information}
\href{./SI.pdf}{Supplementary Information} is available for supplementary information. 
	
\section*{Conflict of Interest}
The authors declare no conflict of interest.
	
\section*{Acknowledgments}
This work was supported in part by NIH Grant R01AI164266 and the Georgia Research Alliance. 

\section*{AI Usage Statement}

The authors used an AI-based language model solely to improve the clarity and grammar of the manuscript. The authors take full responsibility for the study design, mathematical formulations, computational analyses, results, interpretations, and final content of the manuscript.
	
\clearpage
	
\bibliographystyle{unsrt}
\bibliography{ref}

\end{document}